\documentclass{article}
\usepackage{spconf,amsmath,graphicx}

\def\onedot{.}
\def\eg{\emph{e.g}\onedot} 

\def\ie{\emph{i.e}\onedot}

\def \x {\mathbf{x}}

\def \y {\mathbf{y}}

\usepackage{multirow}
\usepackage{multicol}
\usepackage{microtype}

\usepackage{bm}

\usepackage{amsmath}
\usepackage{amssymb}
\usepackage{amsfonts}
\usepackage{algorithm}
\usepackage{algpseudocode}

\usepackage{subfig}

\usepackage{color}

\usepackage{amsthm}

\def\x{{\mathbf x}}

\title{Multi-view Fusion Transformer for Sensor-based Human Activity Recognition}
\name{Yimu Wang, Kun Yu, Yan Wang and Hui Xue}
\address{Alibaba Group}
\begin{document}
%\ninept
%
\maketitle
\begin{abstract}
As a fundamental problem in ubiquitous computing and machine learning, sensor-based human activity recognition (HAR) has drawn extensive attention and made great progress recent years. HAR aims to recognize human activities based on the availability of rich time-series data collected from multi-modal sensors such as accelerometers and gyroscopes. However, recent deep learning methods are focusing on one view of the data, \ie, the temporal view, while shallow methods tend to utilize the hand-craft features for recognition, \eg, the statistics view. 
In this paper, to extract a better feature for advancing the performance, we propose a novel method, namely multi-view fusion transformer (MVFT) along with a novel attention mechanism. First, MVFT encodes three views of information, \ie, the temporal, frequent, and statistical views to generate multi-view features. Second, the novel attention mechanism uncovers inner- and cross- view clues to catalyze mutual interactions between three views for detailed relation modeling. Moreover, extensive experiments on two datasets illustrate the superiority of our methods over several state-of-the-art methods.

\end{abstract}
\begin{keywords}
Multi-view Learning, Sensor, Attention
\end{keywords}
\section{Introduction}
\label{sec:intro}

Understanding human activities is an important and challenging problem in machine learning, which is the basis for healthcare applications, such as rehabilitation, gait analysis, falls detection and falls prevention. Human activity recognition (HAR)~\cite{ICASSP01}, as an effective method towards understanding human activities, has drawn more and more attention recently. The goal of HAR is to classify the activities with RGB image data~\cite{ruiz3DCNNsDistance2017}, WIFI signal~\cite{liTwoStreamConvolutionAugmented2021,WangCYX21} or wearable sensor data~\cite{abedinSparseSenseHumanActivity2019} collected from participants via computational systems. As image data is harder to obtain than wearable sensor data, in this paper, we focus on wearable sensor-based HAR. Specifically, the raw (sensor-based) data is a stream of sensor readings received from limited wearable sensors (typically $4 \times 3$), apart from image data, which is abundant in information with more dimensions (usually $3\times224\times224$). 

Existing HAR methods can be roughly catergorized into two groups, shallow methods and deep methods. Shallow methods generally use PCA, LDA, Fourier transformation and handcrafted features, \eg, mean, variance, median, maximum, minimum, to obtain features, which means the feature extraction procedure is independent of the following classification procedure. That may significantly affect the performance, because these features are not optimal for the following procedure. Recently, as deep learning~\cite{heDeepResidualLearning2016} has shown its superiority of representation learning in various applications, such as image recognition~\cite{simonyanVeryDeepConvolutional2014,wangAdversarialDomainAdaptation2020} and information retrieval~\cite{wangSearchingPrivatelyImperceptible2020,wangPiecewiseHashingDeep2020}, more and more deep HAR methods have been proposed. Deep methods~\cite{abedinSparseSenseHumanActivity2019,ordonezDeepConvolutionalLSTM2016,xiDeepDilatedConvolution2018} capture highly implicit feature and thus achieve better performance than shallow methods leverage the power of deep learning by integrating feature learning and classification into a single framework.

However, while existing deep methods shed their lights and achieve better performances than shallow methods, most of the deep methods only focus on one view of data, \ie, the temporal view (raw data). Deep feature extraction also has its own limits, as it is hard to count how many peaks show in the raw data or what are the mean, maximum and minimum values of data. On the other hand, frequency domain is important to sequential data analysis~\cite{jiangHumanActivityRecognition2015} as different sensor sequences generally have unique periodicity and different amplitude, which are more suitable to be described by frequency domain features, and statistics of data is also able to illustrate some aforementioned criteria, which provide neural networks more views of data and thus may boost the performance.

In this paper, to address this issue, we propose a novel multi-view fusion transformer (MVFT) and an attention mechanism. Specifically, to utilize multi-view data, MVFT encodes them separately and then employs an multi-view fusion attention to explores inter- and cross- view correlations among three views features. First, to learn separate suitable features, MVFT employs a separate transformer block~\cite{vaswaniAttentionAllYou2017} to encode each view. Then, a multi-view fusion attention module is used to enhance mutual cross-view communications and learn discriminative features. Moreover, we conduct extensive experiment on two datasets illustrating the superiority of our methods over several state-of-the-art methods.

\section{Related Work}
\label{sec:rw}

Human Activity Recognition (HAR) is an important and challenging topic in artificial intelligence. The availability of cheap wearable sensors and the advance of computing technology enable low cost, continuous and non-invasive mobile sensing~\cite{GRAVINA201768}. 
Traditional sensor-based HAR systems manually extract a set of features from time-series raw sensor signals and then employs different machine learning algorithms to map features to various classes (human activities), \eg, decision tree~\cite{vankasterenAccurateActivityRecognition2008}, KNN~\cite{Hasan2016HumanAR}, SVM~\cite{bullingRecognitionVisualMemory2011,Anguita2013APD}.
\cite{bullingRecognitionVisualMemory2011} extracts features from data collected from accelerometer and gyroscope, and applies a multi-class SVM to classify six different activities. But, the feature extraction procedure is man-made, handcrafted and empirical, which is not optimal for the following procedure and blocks the way to better performance and end-to-end systems.

Recently, inspired by the advance in deep learning, more and more deep HAR methods are proposed. In contrast to shallow supervised HAR methods, deep HAR methods integrates the feature learning and prediction (classification) procedure and in this way, feature learning can extract better features suitable for the classification procedure. Besides, it offers a powerful ability of latent feature representation, such as specific variance of signals at different scales that reflects the salient pattern of signals. 
Specifically, \cite{yangDeepConvolutionalNeural2015} automates feature learning from the multichannel time series data for HAR task using 2dCNN network, which mainly employs convolution and pooling operations to capture the salient patterns. 
Following that, \cite{xuHumanActivityRecognition2018} utilizes a 1dCNN network with triaxial accelerometer data collected from users' smartphones to recognize three human activities. 
Besides, they found that the activity recognition performance was improved if the input vector dimension was increased. \cite{ordonezDeepConvolutionalLSTM2016} proposes a generic deep framework (DeepConvLSTM) based on convolutional and LSTM recurrent units, which performs sensor fusion naturally and explicitly models the temporal dynamics of feature activation. \cite{inoueDeepRecurrentNeural2016,hammerlaDeepConvolutionalRecurrent2016} explore the performance of LSTM across three representative datasets that contain movement data captured with wearable sensors. \cite{zhaoLearningMonitorMachine2017} employs BiLSTM to encode temporal dependencies and model sequential structure, past and present contextual information. 

However, most the previous HAR methods only use one view of data to generate features, such as temporal view~\cite{abedinSparseSenseHumanActivity2019}, the frequency view~\cite{jiangHumanActivityRecognition2015,laputSensingFineGrainedHand2019} and the statistic view~\cite{qianNovelDistributionEmbeddedNeural2019}. Frequent view helps model to capture frequency features, while statistical feature, \eg, mean, variance and the number of peaks, is far more interpretable but requiring domain knowledge. Three different views of data provide different angles to the data. 
In this paper, to incorporate three views and boost the interactions among three views data, we propose a transformer-like model (MVFT) and a novel multi-view fusion attention mechanism. Comparing with previous methods, MVFT encodes three views of information separately and then uncovers inner- and cross- view clues to catalyze mutual interactions between three views for detailed relation modeling, which further advances the accuracy.

\section{Proposed Methods}
\label{sec:main}

%---------------------------------------
\begin{figure*}[t]
	\begin{center}
		\includegraphics[width=0.9\linewidth]{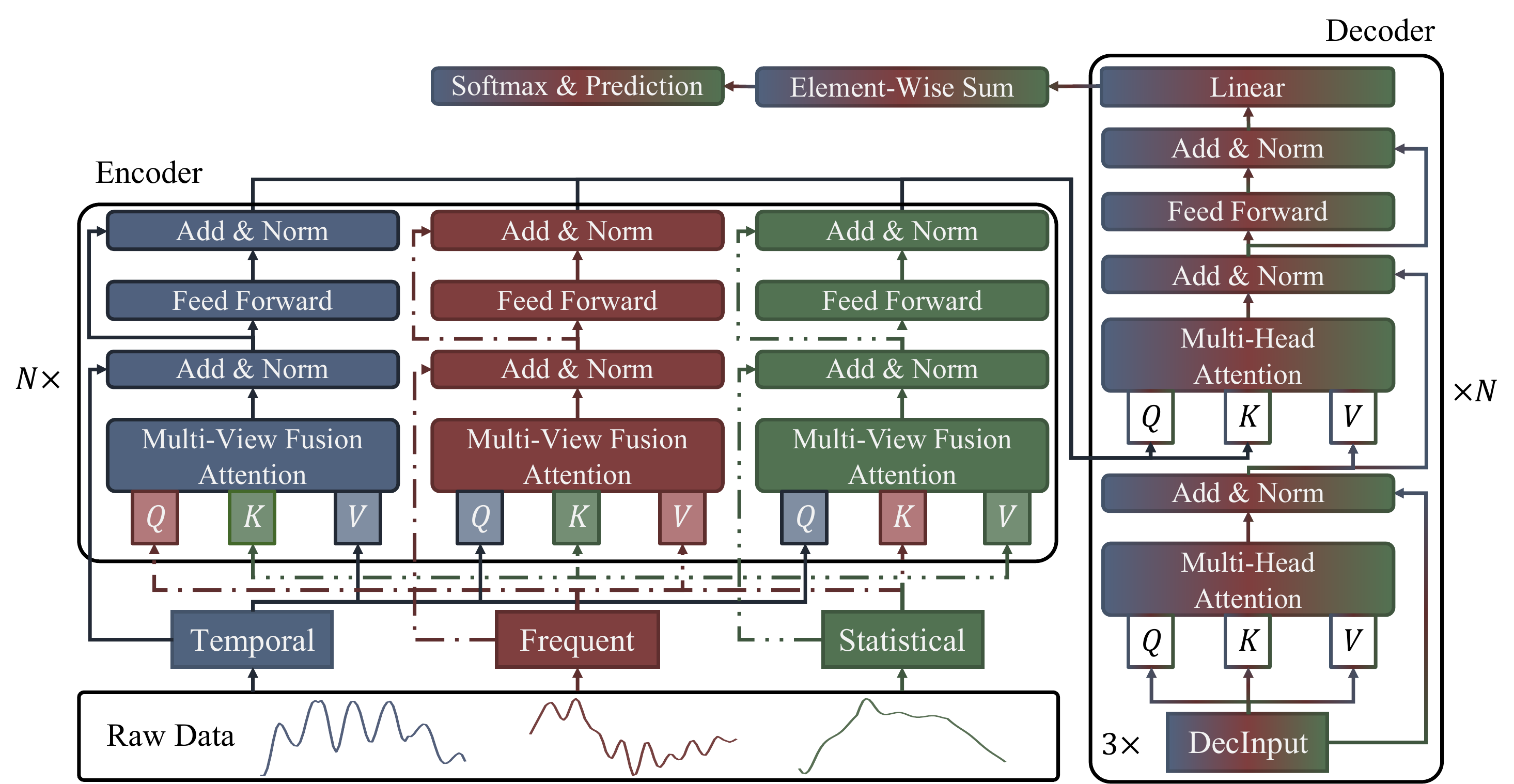}
		\vspace{-0.5em} 
		\caption{
			The architecture of our model. Raw data collected from wearable sensors is firstly transformed to produce three views of data. After that, input sequence (temporal frequent and statistic views) will go through several encoder blocks, which contains a novel multi-view fusion attention to uncover inner- and cross- view clues and catalyze mutual interactions between three views for detailed relation modeling. Following that, the decoder consisting several blocks takes decoder inputs and the outputs of encoder to generate final results with an element-wise sum and a softmax operation. ``DecInput'' refers to the input of decoder and ``temporal'', ``frequent'' and ``statistic'' represent temporal, temporal and statistic views.
		}
		\label{fig:intro}
	\end{center}
\end{figure*}·

\subsection{Problem Formulation and Notation}

In sensor-based HAR, raw input $\x_r = [\x_{r, 1}, \ldots, \x_{r, T}] \in \mathbb{R}^{T\times (dm)}$ is a data-stream of raw time-series samples collected from sensors, where $d$, $m$ and $T$ is the dimension of each sensor, the number of sensors and the total length of the sequence. Without loss of generality, all the sensors are assumed to have the same sampling rate. The corresponding label is denoted as a one-hot vector $\y \in \{0, 1\}^{n_{class}}$, where $n_{class}$ is the number of classes. The goal of HAR is to learn a classifier $f$ using the training data minimizing the empirical error.

\subsection{Our Proposed Methods}

In this part, we elaborately present the data preparation and the proposed multi-view fusion transformer (MVFT) as shown in Fig.~\ref{fig:intro}.

\subsubsection{Data Preparation and Embedding}

Before uncovering inner- and cross- view clues, different views of data should be obtained first. In our methods, three views of data are used, \ie, the temporal, frequent and statistic views, which are shown below,
\begin{itemize}
    \item[1] Temporal view: Temporal view (time domain) provides the raw 
    the response of a system consisting of several sensors, and it illustrates the raw data without any modification and analyzing. 
    \item[2] Frequent view: Frequent view shows the data from a different perspective. It transforms the raw data to its frequency representation, which is known as the "spectrum" of frequency components. An example of such transformation is a Fourier transform. The Fourier transform converts the temporal data into a set of sine waves that represent different frequencies. In this paper, we use it to obtain the Frequent view. % of data.
    \item[3] Statistic view: Statistic view of data is a list of characteristics, such as the maximum, minimum, average value and number of peaks, extracted from time series. It depicts raw data (temporal view of data) from different perspectives.
\end{itemize}

Each input sequence starts with a special token ``[CLS]'', \ie, $\x = [\text{``[CLS]''}, \x_r,\text{``[CLS]''},\x_f, \text{``[CLS]''}, \x_s]$, where $\x_f$ and $\x_s$ are the frequent and statistic views of data. 

Besides, to explicitly transfer position, time and three views information to MVFT, we use five types of embedding, \ie, position embedding, time embedding, temporal embedding, frequent embedding and statistic embedding\footnote{We do not employ a segment embedding indicating the three views of data because we believe that it is unnecessary as MVFT uses three separate Transformer-like module to extract features.}. \textbf{Position embedding: } Following~\cite{vaswaniAttentionAllYou2017}, MVFT incorporates a learnable position embedding to every input in the sequence to learn the correlation with data and position, as self-attention-like mechanism does not consider the order of input data. \textbf{Time embedding: } Different from other tasks~\cite{zhuActBERTLearningGlobalLocal2020,dosovitskiyImageWorth16x162021}, time is significantly important in HAR as it provides us the delay between any two elements, and it enables models to extract features even when the time gap between any two elements varies. \textbf{temporal embedding}, \textbf{frequent embedding} and \textbf{statistic embedding} are the corresponding segment embeddings of three views of data.

The decoder of transformer-like models needs inputs too. We use vectors with all elements being $1$ as the input of decoders.

\subsubsection{Multi-View Fusion Transformer}

We design a Multi-View Fusion Transformer (MVFT) to better encode three sources of information, \ie, the temporal view, the frequent view and the statistic view as shown in Fig.~\ref{fig:intro}. 

MVFT employs three transformers to encode three view features equally instead of using only one transformer, which enables MVFT to separately learn feature for three views. To promote the flow of information between different views, we propose a multi-view fusion attention mechanism, which incorporates three views information and enhance the interactions between them. With cross-view interactions, MVFT can dynamically focus on judicious cues for further prediction. Specifically, the multi-view fusion attention mechanism takes \emph{three different views features as query, key and value} and then performs multi-head attention on them. The multi-view attention is as follows,
\begin{equation}
\begin{aligned}
    \operatorname{MultiView} (Q, K, V) & = \operatorname{Concat}_{h \in [H]} (\text{head}_h) W^O \,,\\
    \text{head}_h &= \operatorname{Attention}(QW_{h}^{Q}, KW_{h}^{K}, VW_{h}^{k})\,, \\
    \operatorname{Attention}(Q, K , V) & = \operatorname{softmax}\left(\frac{QK^\top}{\sqrt{d_k}}\right) V\,,
\end{aligned}
\end{equation}
where $d_k$ is the dimension of query, $\operatorname{softmax}(\cdot)$ is the softmax operation, $Q$, $K$ and $V$ are query, key and value, $\operatorname{Concat}(\cdot)$ is the concatenation operation, $W_{*}^{Q}, W_{*}^{K} \in \mathbb{R}^{d_{model} \times d_{k}}, W_{*}^{V} \in \mathbb{R}^{d_{model} \times d_{v}}$ are three learnable projections, $H$ is the number of head. With this multi-view fusion attention, extensive interactions can be extracted from three views features and further improve final performance.

MVFT only employs multi-view fusion attention in the encoder as the task of encoder is to learn powerful and discriminative representations, while the goal of decoder is to advance the interaction between the learnt three views features generated by the encoder and the corresponding decoder input.

\section{Experiments}
In this section, we carry out experiments to empirically evaluate the performance of MVFT on two wearable sensor benchmarks, and then compare it to state-of-the-art approaches.

% Please add the following required packages to your document preamble:
% \usepackage{multirow}
\begin{table}[t!]
\begin{tabular}{|l|c|c|c|}
\hline
\multirow{2}{*}{Method} & \multicolumn{3}{c|}{Acc (\%)} \\ \cline{2-4} 
                        & Wisdom   & CRO\_30  & CRO\_50 \\ \hline
CNN1d                   & 87.25  & 82.87  & 82.18 \\ \hline
CNN2d                   & 89.56  & 83.40  & 82.25 \\ \hline
BiLSTM                  & 44.67  & 82.35  & 82.78 \\ \hline
DeepConvLSTM            & 32.20  & 91.02  & 90.96 \\ \hline
LSTM                    & 37.72  & 82.27  & 89.42 \\ \hline
RNN                     & 30.96  & 92.35  & 91.21 \\ \hline
SparseSense             & 78.54  & 88.30  & 88.61 \\ \hline\hline
Transformer             & 91.53  & 91.29  & 91.11 \\ \hline
Ours                    & \textbf{96.71}  & \textbf{99.40}  & \textbf{99.18} \\ \hline
\end{tabular}
\vspace{-0.5em}
\caption{Comparison of Accuracy (\%) on three datasets, \emph{Wisdom}, \emph{CRO\_30} and \emph{CRO\_50}. Best in bold.}\label{tab:main}

\end{table}

\subsection{Datasets and Details}
Two datasets, Wisdom~\cite{ActivityRecognitionUsing} and a collected dataset named CRO are used for evaluation. 
\textbf{Wisdom} is collected using accelerometers built into phones~\cite{ActivityRecognitionUsing}. Subjects carried the Android phone in their front pants leg pocket and were asked to perform six regular activities, \ie, walking, jogging, ascending stairs, descending stairs, sitting and standing. The sequence length of Wisdom is $30$.
\textbf{CRO} is collected by four sensors on smartphones and manually labeled into two classes. The smartphones are embedded with accelerometers, magnetometers and gyroscopes. We collect 89874 raw data and form two datasets, \ie, $\text{CRO}\_{30}$ and $\text{CRO}\_{50}$. The former is the dataset with length of 30 while the length of latter is $50$. 

MVFT is implemented based on PyTorch~\cite{paszkePyTorchImperativeStyle2019} with eight NVIDIA P100 GPUs and optimized by the mini-batch Adam with the size of $256$. The learning rate is initialized as 0.0001. We compare our method with several state-of-the-art (SOTA) methods, \ie, CNN1d~\cite{xuHumanActivityRecognition2018}, CNN2d~\cite{yangDeepConvolutionalNeural2015}, BiLSTM~\cite{zhaoLearningMonitorMachine2017}, DeepConvLSTM~\cite{ordonezDeepConvolutionalLSTM2016}, LSTM~\cite{inoueDeepRecurrentNeural2016,hammerlaDeepConvolutionalRecurrent2016}, RNN~\cite{zarembaRecurrentNeuralNetwork2015}, SparseSense~\cite{abedinSparseSenseHumanActivity2019} and Transformer~\cite{vaswaniAttentionAllYou2017}.

\subsection{Main Results}

The results on two datasets are presented in Table~\ref{tab:main}. On all datasets, our proposed MVFT outperforms all the methods with the power of multi-view fusion attention. Specifically, on Wisdom, MVFT outperforms CNN2d, the best SOTA, and Transformer by 7.15\% and 5.18\%, respectively, while on CRO\_30, MVFT outperforms DeepConvLSTM and RNN, the best SOTAs, and Transformer by 8.38\% and 8.11\%. On CRO\_50, the gaps between MVFT and DeepConvLSTM, LSTM, RNN and Transformer are extended to 8.22\%, 8.22\%, 8.22\% and 8.07\%.

\vspace{-0.5em}

% Please add the following required packages to your document preamble:
% \usepackage{multirow}
\begin{center}
\begin{table}[t!]
\resizebox{\columnwidth}{27mm}{
\begin{tabular}{|l|c|c|c|c|c|c|}
\hline
\multirow{2}{*}{Method}     & \multicolumn{3}{c|}{Views}            & \multicolumn{3}{c|}{Acc (\%)} \\ \cline{2-7} 
                             & T   & F   & S & Wisdom  & CRO\_30  & CRO\_50  \\ \hline\hline
\multirow{7}{*}{Transformer} & \checkmark &            &             & 91.53   & 91.29    & 91.11    \\ \cline{2-7} 
                             &            & \checkmark &             & 85.22   & 98.14    & 95.92    \\ \cline{2-7} 
                             &            &            & \checkmark  & 93.29   & 95.21    & 95.78    \\ \cline{2-7} 
                             & \checkmark & \checkmark &             & 94.26   & 97.17    & 96.15    \\ \cline{2-7} 
                             & \checkmark &            & \checkmark  & 95.90   & \textbf{98.61}    & \textbf{97.82}    \\ \cline{2-7} 
                             &            & \checkmark & \checkmark  & 95.30   & 98.46    & \textbf{97.82}    \\ \cline{2-7} 
                             & \checkmark & \checkmark & \checkmark  & \textbf{96.21}   & 97.93    & 96.38    \\ \hline \hline
\multirow{4}{*}{Ours}        & \checkmark & \checkmark &             & 96.02   & 98.10    & 98.20    \\ \cline{2-7} 
                             & \checkmark &            & \checkmark  & 95.60   & 98.79    & 98.15    \\ \cline{2-7} 
                             &            & \checkmark &       \checkmark       & 96.66   & 98.86    & 99.10    \\ \cline{2-7} 
                             & \checkmark & \checkmark & \checkmark  & \textbf{96.71}   & \textbf{99.40}    & \textbf{99.18}    \\ \hline
\end{tabular}
}
\vspace{-0.5em}
\caption{Comparison of Accuracy (\%) on three datasets, \emph{Wisdom}, \emph{CRO\_30} and \emph{CRO\_50}. Best in bold.}\label{tab:ablation}
\end{table}
\end{center}

\subsection{Ablation Study}

To evaluate how three views weigh in the performance, we conduct ablation study shown in Table~\ref{tab:ablation} using MVFT and Transformer. In the most of results, using more views means better performance, while in some scenarios, more views could be harmful to performance, especially in the experiments with Transformer. 
Using any two views in Transformer can improve the accuracy comparing with the model using only one view on all the datasets, while adding one more view may cause the performance drop. 
While in the experiments using MVFT, on all three datasets, incorporating more views can gain performance boost. Specifically, on Wisdom, the model using three views outperforms the models using any two views by 0.69\%, 1.11\% and 0.05\%. Similar results can be seen on others.

\section{Conclusion}

In this paper, we presented a novel cross-view fusion transformer methods named MVFT. MVFT was the HAR deep method that learned powerful and discriminative multi-view features. The key contribution was incorporating and boosting the inner- and inter- interactions thus enabling one view features absorbing other views' information to further focus on general similar features across different views. 
Extensive experiments showed the superiority of our method over several state-of-the-art methods. In the future, we would like to explore the potential of transformer on extremely long sequential data with multi-view fusion attention. 
% Besides, as transformer-like architecture played an important role in the computer vision area, 

% References should be produced using the bibtex program from suitable
% BiBTeX files (here: strings, refs, manuals). The IEEEbib.bst bibliography
% style file from IEEE produces unsorted bibliography list.
% -------------------------------------------------------------------------
\begin{small}
\bibliographystyle{IEEEbib}
\bibliography{ref}
\end{small}
\end{document}